\documentclass[%
reprint,
superscriptaddress,
amsmath,amssymb,
aps,
]{revtex4-2}

\usepackage{graphicx}
\usepackage{dcolumn}
\usepackage{bm}
\usepackage{siunitx}
\usepackage[version=4]{mhchem}
\usepackage{multirow}
\usepackage{tikz}
\usepackage{hyperref}
\usepackage{url}

\begin{document}

\title{\texttt{AIM$^2$DAT}: A Python-based Automated Ab Initio Material Modeling and Data Analysis Toolkit}

\author{Holger-Dietrich Saßnick}
\email{holger-dietrich.sassnick@umontpellier.fr}
\affiliation{Carl von Ossietzky Universit\"at Oldenburg, Physics Department, 26129 Oldenburg, Germany}
\affiliation{ICGM, Université de Montpellier, CNRS, ENSCM, 34293 Montpellier, France}

\author{Joshua Edzards}
\affiliation{Carl von Ossietzky Universit\"at Oldenburg, Physics Department, 26129 Oldenburg, Germany}
\affiliation{Friedrich-Schiller-Universit\"at Jena, Institut f\"ur Festk\"orpertheorie und -Optik, 07743 Jena, Germany}

\author{Timo Reents}
\affiliation{Carl von Ossietzky Universit\"at Oldenburg, Physics Department, 26129 Oldenburg, Germany}
\affiliation{PSI Center for Scientific Computing, Theory and Data, Paul Scherrer Institute, 5232 Villigen PSI, Switzerland}

\author{Caterina Cocchi}
\affiliation{Carl von Ossietzky Universit\"at Oldenburg, Physics Department, 26129 Oldenburg, Germany}
\affiliation{Friedrich-Schiller-Universit\"at Jena, Institut f\"ur Festk\"orpertheorie und -Optik, 07743 Jena, Germany}

\date{\today}


\begin{abstract}
The emergence of data-driven computational materials science offers unprecedented opportunities to explore complex material landscapes, complementing experimental research with the discovery of novel compounds.
To enable these developments, it is essential to establish robust, reliable, and easy-to-use software supporting workflow automation and large dataset processing.
Herein, we introduce the Automated Ab Initio Materials Modeling and Data Analysis Toolkit (\texttt{aim$^2$dat}), a \texttt{Python} package offering a user-friendly interface to generate and handle big data, design high-throughput workflows based on density functional theory calculations, and analyze the output.
Its key features include interfaces to online databases for structure query and analysis, high-throughput screening routines, and seamless integration of machine learning models.
The capabilities of \texttt{aim$^2$dat} are showcased with a variety of use-cases, ranging from photocathode materials to metal-organic frameworks.
\end{abstract}

\maketitle

\section{Introduction}

Computational data-driven studies and high-throughput screening have matured into valuable tools, complementing experimental efforts to discover novel materials and unravel structure-property relationships for key applications, such as energy storage~\cite{chen+15jpcl,bena+22aem}, carbon dioxide direct air capture~\cite{lim+25matter}, solar cells~\cite{wu+25sci}, and catalysis~\cite{rose+19jcc, tran+23acsc}.
Density functional theory (DFT) has emerged as the workhorse \textit{ab initio} method for treating molecules and crystals without empirical parameters and at viable computational costs~\cite{hafn+06mrsb}. Nowadays, DFT is routinely embedded in workflow engines~\cite{hube+20sd, uhri+21cms, fireworks, atomate, pyiron, simstack} to scale up the number of calculations, generating the large datasets required to unravel structure-property relationships that are essential for applications or train machine learning (ML) models~\cite{schl+19jpm, schm+19npjcm, west+21jcp, xu+23mgea, nyan25nm, ghar+26sd, hube+26dd}.

Despite these achievements, the reproducibility and transferability of high-throughput workflows remain challenging~\cite{hegd+23prm, tawf25jcim, boso+25natrp, jans+25dd}.
Discrepancies in data provenance and the inherent complexity of managing multi-step simulations may lead to silent errors that compromise data veracity, ultimately limiting the predictive reliability of ML models~\cite{hima+19as, reev-moth26jmca}. To fully realize the potential of the artificial intelligence (AI) era, the community must transition from mere scalability to a framework of FAIR-compliant (Findable, Accessible, Interoperable, and Reusable) automation~\cite{wilk+16sd,sche+22nat}. It is therefore imperative to develop integrated toolkits that ensure not only efficient high-throughput simulations but also the high-fidelity datasets required to navigate increasingly complex and multidimensional material landscapes~\cite{bai-zhan25nml}.

To address these challenges, we introduce the \textit{\textbf{A}utomated Ab \textbf{I}nitio \textbf{M}aterials \textbf{M}odeling and \textbf{D}ata \textbf{A}nalysis \textbf{T}oolkit} (\texttt{aim$^2$dat}) software package, a \texttt{Python}-based infrastructure designed to streamline the production and analysis of high-quality DFT datasets. This package is built on modular design principles, ensuring that core components interlock seamlessly to manage complex computational workflows. Here, we demonstrate the versatility of \texttt{aim$^2$dat} through a series of diverse use-cases, ranging from inorganic semiconductors used for modern photocathodes in particle accelerators and their surfaces, to metal-organic frameworks (MOFs). Beyond discussing the capability of \texttt{aim$^2$dat} to straightforwardly reveal the structure-property relations in these diverse material classes, we present the successful integration of ML models for crystal structure prediction.

This paper is organized as follows: In Sec.~\ref{sec:features}, we introduce the design principles, discussing in detail aspects related to data mining, generation, and curation, the high-throughput screening workflow, and the tools for data analysis and ML integration. In Sec.~\ref{sec:use_cases}, we showcase the capabilities of \texttt{aim$^2$dat} with a few relevant examples from our recent research: semiconducting photocathodes, metal-organic frameworks, and other applications in the areas of organic semiconductors and ternary transition-metal oxides. In Sec.~\ref{sec:conclusions}, we summarize the main features of \texttt{aim$^2$dat} and draw our conclusions and outlook.

\section{Features and Design Principles}\label{sec:features}

\begin{figure*}
	\centering
    \includegraphics[width=\linewidth]{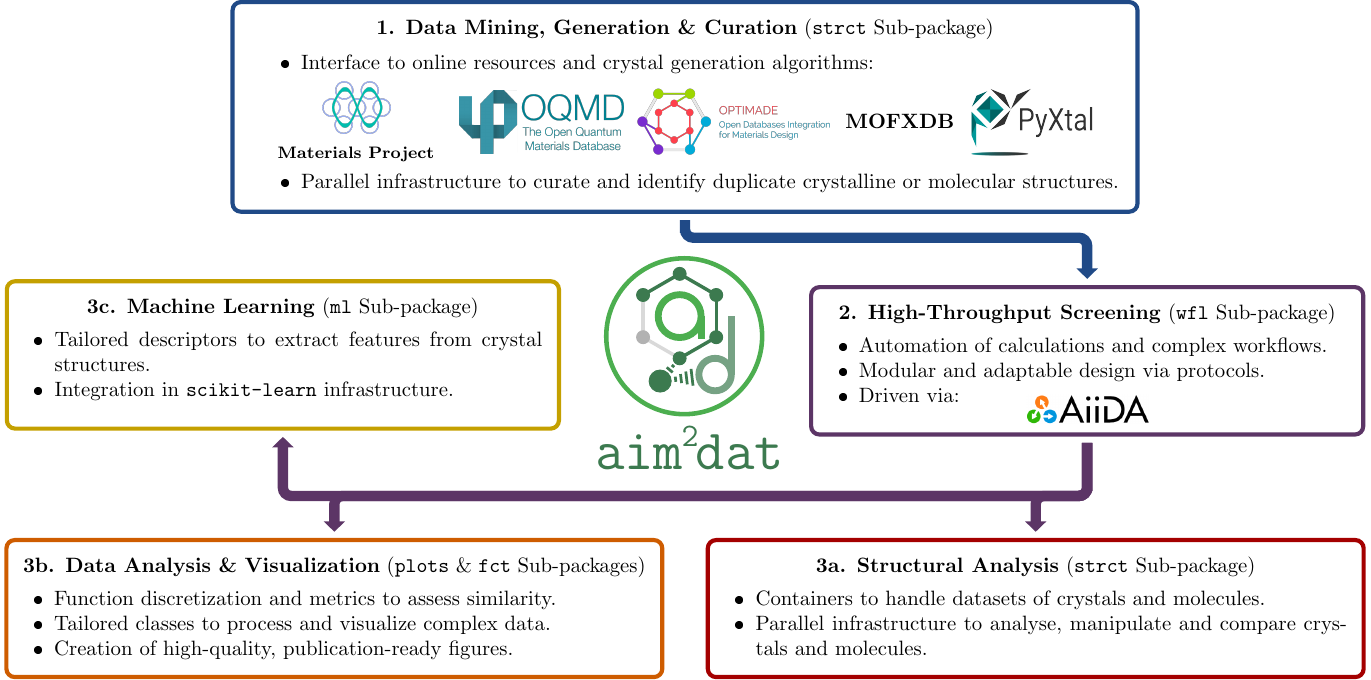}
    \caption{\label{fig:htp_workflow} Schematic representation of the \texttt{aim$^2$dat} architecture, illustrating the integrated pipeline from initial data mining and structural generation to automated simulation management and machine-learning-assisted analysis. Individual modules are depicted as framed units, highlighting the specialized classes and third-party interfaces provided by the package.}
\end{figure*}

The main purpose of \texttt{aim$^2$dat} is to provide a streamlined framework for managing \textit{ab initio} high-throughput studies.
Developed as an open-source \texttt{Python} library, this package is distributed free of charge under the LGPL-2.1 license with the source code and comprehensive documentation being hosted on \texttt{GitHub}~\cite{repository, docs}.
\texttt{aim$^2$dat} offers a modular toolkit, enabling user-designed ``recipes'' optimized for specific research objectives. In the following subsections, we present in detail the core features of this package, which is structured around the three stages of high-throughput workflow: structure generation, calculation management, and data analysis. A schematic visualization of the structure of \texttt{aim$^2$dat} and the interconnections among its parts is provided in Fig.~\ref{fig:htp_workflow}.

\subsection{Data Mining, Generation, and Curation}
The \texttt{strct} sub-package provides a robust toolset to construct and curate the input structural pools for \textit{high-throughput} simulations.
Given the wealth of open-access repositories, \texttt{aim$^2$dat} streamlines the initial data mining stage via the \texttt{StructureImporter} class. This interface allows users to query online resources such as the Open Quantum Materials Database (OQMD)~\cite{saal+13jom}, Materials Project (MP)~\cite{jain+13aplm}, and MOFXDB~\cite{mofxdb}, using specified compositions. Integration with the \texttt{optimade}~\cite{ande+21sd} API further extends this access to repositories such as Materials Cloud~\cite{tali+20sd}, NOMAD~\cite{nomad}, and AFLOWlib~\cite{curt+12cms}. Furthermore, for crystal structure prediction (CSP) workflows, \texttt{aim$^2$dat} supports the stochastic generation of new candidates using the \texttt{PyXtaL} package~\cite{pyxtal} as a backend.

The fundamental data unit is the \texttt{Structure} object, which encapsulates all structural and metadata information, including atomic positions, unit cell vectors, and periodic boundary conditions. To ensure interoperability across the computational materials science ecosystem, this class offers seamless conversion to and from \texttt{ase}~\cite{ase}, \texttt{pymatgen}~\cite{pymatgen}, \texttt{AiiDA}~\cite{hube+20sd}, and \texttt{openmm}~\cite{openmm} objects. Scaling these units for high-throughput studies is managed by the \texttt{StructureCollection} and \texttt{StructureOperations} classes. The former acts as a centralized container for database storage (\texttt{AiiDA}) or file-based serialization (\texttt{HDF5}), and can be exported as a \texttt{pandas} \texttt{DataFrame} for rapid data exploration and attribute filtering.

To abstract away manual looping, the \texttt{StructureOperations} class enables the parallel execution of manipulation or analysis methods across the entire collection. Key functionalities include:
\begin{itemize}
    \item \textbf{Pipeline Concatenation and Customization:} Users can construct complex workflows to extend input structure pools by performing chemical substitutions or appending functional groups. This process can be conveniently handled in a high-throughput fashion~\cite{sass+24ic, edza+24jcp} as extensively discussed in Sec~\ref{example:mofs}.
    \item \textbf{Automated Filtering:} To optimize computational efficiency, \texttt{aim$^2$dat} includes routines such as the F-fingerprint method~\cite{ogan-vall09jcp} to identify and remove duplicate structures. This ensures that the finalized structural pool is non-redundant and computationally lean before being dispatched to the high-throughput simulation engines described in the following sections.
\end{itemize}

By integrating these generation and curation routines, \texttt{aim$^2$dat} ensures that structural ensembles are both chemically diverse and computationally non-redundant. This streamlined data preparation serves as a foundation for the high-throughput simulation engines described in the following section.

\subsection{High-Throughput Screening}

Once the structure pool is established, \texttt{aim$^2$dat} orchestrates automated workflows on each structure to perform the underlying DFT calculations.
These high-throughput capabilities are powered by the \texttt{AiiDA} infrastructure~\cite{pizzi+16cms, hube+20sd, uhri+21cms}, which manages workflow execution and maintains full data provenance within its underlying database. \texttt{aim$^2$dat} offers a suite of custom \texttt{AiiDA} \texttt{WorkChain} classes categorized into \textit{core} tasks (e.g., unit cell relaxation or electronic band structure  calculation), and \textit{combined} tasks for more complex operations like surface slab convergence (see Sec.~\ref{example:pc}).
A high level of automation is achieved by utilizing the workflow engine of the AiiDA software package: once a task is triggered by the user, all necessary DFT calculations are automatically started, monitored, and the output files are parsed and stored in the underlying database for future analysis without human intervention.

The current version of \texttt{aim$^2$dat} is optimized for DFT calculations performed with the \texttt{CP2K} code~\cite{kueh+20jcp}; other codes can be used, provided their interface with \texttt{AiiDA}. External packages like \texttt{enumlib}~\cite{hart+12cms}, \texttt{critic2}~\cite{oter+14cpc}, and \texttt{chargemol}~\cite{manz-lima16rscadv} are integrated for specialized pre- and post-processing. To maximize the success rate of large-scale screenings, each \texttt{WorkChain} includes automated error-handling routines that dynamically adjust parameters, such as the Kohn-Sham mixing scheme or geometry relaxation tolerances, to overcome common convergence hurdles~\cite{sass-cocc22jcp}.

To enhance  flexibility beyond standard routines, \texttt{aim$^2$dat} introduces a workflow infrastructure consisting of the \texttt{WorkflowBuilder} and \texttt{MultipleWorkflowBuilder} classes. These tools enable a modular workflow design through the definition of a \textit{protocol}, i.e., a configuration that encompasses input parameters, task dependencies, and the specific results to be extracted.  Users can utilize pre-defined protocols for common applications or define custom ones via \texttt{Python} dictionaries or \texttt{YAML} files.

Each workflow instance is constructed from distinct input data nodes, stored in the \texttt{AiiDA} database, and subdivided into the three categories:
\begin{itemize}
	\item The \textbf{parent node} defines the unique structural input for each workflow instance.
	\item The \textbf{general input nodes} are predefined parameters specified by the chosen \textbf{protocol}.
	\item The \textbf{user input nodes} are custom parameters tailored to the specific use case.
\end{itemize}

Within this framework, every \texttt{AiiDA} process is treated as a \textit{task}. Dependencies are handled by mapping the outputs of the preceding tasks to the input of subsequent ones, allowing complex simulations to be built from modular workflow components. Upon initialization, \texttt{aim$^2$dat} automatically scans the database to match existing processes with the current task requirements, returning a categorized status list of accomplished, failed, or pending tasks.
While the \texttt{WorkflowBuilder} manages a single workflow instance, the \texttt{MultipleWorkflowBuilder} acts as a high-level wrapper to scale these operations across an entire structural ensemble. This allows the same ``recipe'' to be applied to the entire configurational space simultaneously. To facilitate data management, process states and calculated properties are exposed as \texttt{pandas} \texttt{DataFrame} objects~\cite{pandas1, pandas2}, providing a familiar and powerful interface for high-throughput data analysis and post-processing.

\subsection{Data Analysis and Visualization}

A core strength of \texttt{aim$^2$dat} is its ability to streamline the transition from raw simulation output to physical insight.
This library implements a wide array of structural analysis, ranging from basic geometric properties, such as inter-atomic distances, angles, and coordination environments, to complex descriptors such as radial distribution functions and structural order parameters. The toolkit also  features a robust visualization suite interfacing \texttt{matplotlib}~\cite{matplotlib} for publication-ready static graphics and \texttt{plotly}~\cite{plotly} for interactive data exploration. To optimize the user experience, the plotting architecture is decoupled into two stages: data ingestion/processing and rendering. This design allows users to aggregate multiple datasets into a single figure and adjust visual settings without re-processing the underlying data.

Additionally, \texttt{aim$^2$dat} includes a specialized functional analysis module. This module provides tools to compare one-dimensional functions, such as the electronic density of states. Available metrics range from standard distance measures (e.g., Euclidean or cosine similarity) to sophisticated algorithms capable of handling spectral shifts and broadening, such as the non-uniform discretization grid proposed by Kuban \textit{et al.}~\cite{kuba+22sd}, using a non-uniform discretization grid. By providing these integrated analytical tools, \texttt{aim$^2$dat} bridges the gap between large-scale data generation and the extraction of chemically meaningful trends.
Crucially, while DFT codes typically output calculations to a high decimal precision, the physically meaningful number of significant figures depends heavily on the target property or material class under investigation. Acting as a versatile, domain-agnostic infrastructure rather than a rigid, single-purpose workflow, \texttt{aim$^2$dat} deliberately delegates numerical formatting flexibility to the user during the final aggregation and visualization phases. The diverse case studies presented in Sec.~\ref{sec:use_cases} intentionally retain the specific formatting paradigms of their respective original studies to showcase this architectural adaptability.

\subsection{Machine Learning Integration}

To bridge the gap between high-throughput data generation and predictive modeling, \texttt{aim$^2$dat} offers seamless integration with the \texttt{scikit-learn} ecosystem~\cite{scikit-learn}. The package provides so-called descriptors as custom \texttt{Transformer} classes specifically designed to map complex atomic configurations into structured feature vectors that are input to ML model.
These descriptors are tightly integrated with the \texttt{StructureOperations} class, allowing for the parallelized featurization of massive structural ensembles.

Beyond feature extraction, the toolkit simplifies the ML pipeline by providing tailored routines for dataset curation, including stratified splitting methods for training, validation, and test subsets. This infrastructure ensures that the high-quality data generated via automated DFT workflows can be directly and efficiently utilized to train surrogate models. Furthermore, by maintaining native object interoperability with common libraries like \texttt{ase} and \texttt{pymatgen}, this framework seamlessly interfaces with external machine learning interatomic potential pipelines. This streamlines both the automated generation of high-fidelity training datasets, containing reference energies, forces, and stresses, and the deployment of advanced neural network potentials.
By formalizing this connection between structural data and predictive algorithms, \texttt{aim$^2$dat} enables the rapid iteration cycle required for the AI-driven discovery and characterization of novel materials.

\section{Examples and Use Cases}\label{sec:use_cases}

In the following, we demonstrate the versatility of \texttt{aim$^2$dat} across three distinct research applications. The following three selected use cases highlight how the modular architecture of the library enables the seamless management of diverse data-driven studies. Our discussion focuses primarily on software features; readers interested in the detailed scientific findings are referred to the original publications.

The development of \texttt{aim$^2$dat} was initially driven by the study of alkali-based semiconductors for photocathode applications. In Section~\ref{example:pc}, we discuss how the package facilitated the high-throughput screening of both bulk crystals and surfaces~\cite{sass-cocc22jcp,sass-cocc24npjcm}.
Leveraging its modular design, the workflows implemented in \texttt{aim$^2$dat} were subsequently adapted for MOFs. This transition, discussed in Sec.~\ref{example:mofs}, showcases the extensibility of the package when addressing the unique challenges of hybrid organic-inorganic materials, such as different building blocks, large unit cells, and complex bonding environments~\cite{sass+24ic, edza+24jcp}.
In the third example (Sec.~\ref{example:ml}), we highlight the synergy between DFT and ML to devise a novel crystal structure prediction scheme for binary systems~\cite{sass-cocc25ats}. Finally, Sec.~\ref{example:misc} provides a brief overview of additional examples illustrating the versatility of this library across the broader computational materials science landscape.

\subsection{High-Throughput Screening of Semiconducting Photocathode Materials}\label{example:pc}

\begin{figure*}
	\centering
    \includegraphics[width=\linewidth]{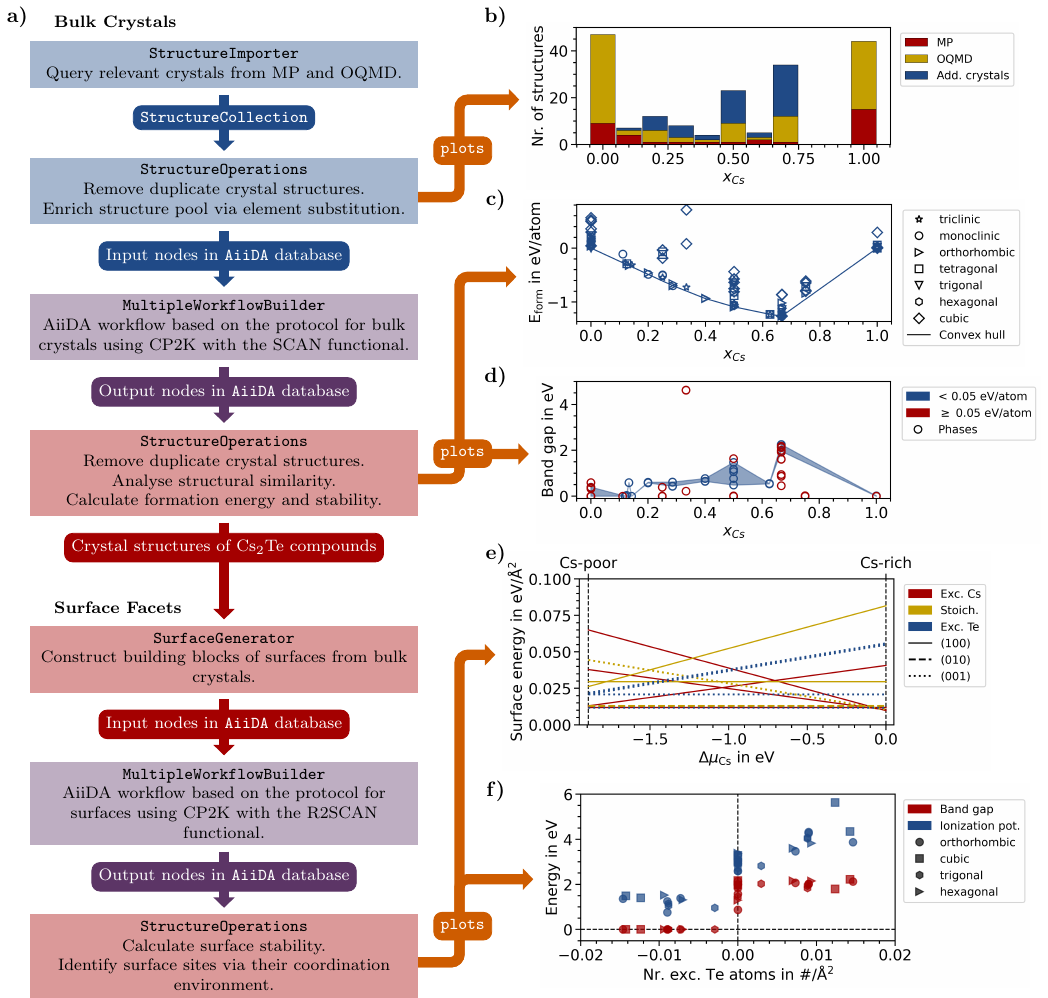}

    \caption{\label{fig:pc_workflow} (a) Schematic representation of the high-throughput workflow with individual tasks (rectangles), and their respective output (rounded boxes). (b) Compositional distribution of the initial structural pool mined from MP (red), OQMD (yellow), and the chemically substituted candidates (blue).
    (c) Formation energies ($E_\mathrm{f}$) with convex hull and (d) band gaps of stable Cs-Te crystals as a function of Cs concentration.
    (e) Surface energies of low-Miller-index Cs$_2$Te facets and (f) corresponding electronic properties (band gap in red and ionization potential in blue) as a function of the excess Te atoms at the surface. Data sourced from Refs.~\cite{pc_bulk_zenodo, pc_surface_zenodo}.}
\end{figure*}

Forming an integral element of photoinjectors, photocathodes are used as electron
sources in particle accelerators.
While metallic cathodes have been state-of-the-art for decades~\cite{cair-sams66josa,down+86apl}, the need for ultrabright electron beams for advanced techniques like electron diffraction~\cite{hada+20acr}, microscopy~\cite{zhu-durr15pt}, and high-resolution spectroscopy~\cite{call+21prep}, has shifted the attention to semiconducting materials. Alkali-antimonides and -tellurides, with their band-gap in the infrared and visible region and low emission thresholds, are ideal candidates for this application~\cite{zhan+17jvstb,scha+23jmcc,moha+23micromachines}. The structural and compositional variability of these material classes, directly inherited by the evaporation techniques adopted for sample growth, calls for reliable computational studies predicting their electronic, optical, transport, and thermal properties~\cite{soua+17pb,cocc+19jpcm,cocc+19sr,amad+21jpcm,sass-cocc21es,cocc-sass21micromachines,zhon+21ijer,shar+23acsaem,sant+24jpm,schi-cocc25prm,bell+25prb,wang+25sr}.

The success of DFT and related methods in predicting and understanding the properties of semiconducting photocathodes~\cite{cocc-sass21micromachines} has stimulated the development of automated schemes, either aiming at discovering the optimal photocathode material across the entire periodic table~\cite{anto+21am} or targeting a medium-sized configuration space restricted to a family of materials. We have chosen the second path focusing on cesium telluride crystalline polymorphs and surfaces~\cite{sass-cocc22jcp,sass-cocc24npjcm}, and (multi-)alkali antimonides~\cite{schi+24ats}. These studies, conducted in the last few years, have boosted the development of \texttt{aim$^2$dat} and have represented a crucial input for the accurate characterization of material properties from many-body perturbation theory~\cite{schi-cocc25prm} and the development of an advanced first-principles approach to simulate photoemission~\cite{schi+26arxiv}.

The workflow implemented for Cs-Te polymorphs is summarized in Fig.~\ref{fig:pc_workflow}a. The starting points are the crystal structures queried from MP and OQMD. The F-Fingerprint method~\cite{ogan-vall09jcp} as implemented in the \texttt{StructureOperations} class is based on discrete radial distribution functions $F_{AB}$ of all element pairs ($A$, $B$) with bin width $d_B$ up to a pre-defined cut-off distance,
\begin{equation}
    F_{AB}(\vec{r}) = \sum_{A_i}^{N_{A,\text{cell}}}\sum_{B_j}^{N_B} \frac{V_{\text{cell}}}{4\pi\,\vec{R}_{ij}^2\, N_A N_B\, d_B} \delta(\vec{r} - \vec{R}_{ij}) - 1,
\end{equation}
enabling filtering out duplicates by using the cosine distance as a similarity metric.
This procedure led to a total of 132 inequivalent polymorphs.

To increase the pool size and the variety of the screened crystals, we additionally selected from the aforementioned databases binary crystals containing chemically similar elements (K or Rb cations and Se or Po anions) and subsequently substituted them with Cs and/or Te. This procedure increased the input pool for high-throughput calculations to 184 binary crystals, stored in a \texttt{StructureCollection} object in \texttt{AiiDA} (Fig.~\ref{fig:pc_workflow}b).

The automated screening was conducted using the \texttt{cp2k-crystal-standard\_v1.0} protocol shipped with \texttt{aim$^2$dat} and implementing the following task sequence:
\begin{itemize}
    \item \texttt{seekpath\_analysis} relies on the \texttt{SeeK-path} library~\cite{seekpath} to obtain the primitive unit cell of the crystal (which is used for all DFT calculations) and the \textbf{k}-path through the Brillouin zone.
    \item \texttt{FindSCFWorkChain} defines appropriate settings to converge the Kohn-Sham equations.
    \item \texttt{CellOptWorkChain} optimizes lattice parameters and atomic positions.
    \item \texttt{BandStructureWorkChain} and \texttt{PDOSWorkChain} compute the electronic band structures and density of states, respectively.
\end{itemize}

A physically relevant cutoff criterion for the high-throughput screening is the energetic stability of the crystals, expressed by their formation energy:
\begin{equation}
	E_\mathrm{f}(\mathrm{Cs}_x\mathrm{Te}_{1-x}) = E(\mathrm{Cs}_x\mathrm{Te}_{1-x})
	- [xE(\mathrm{Cs}) + (1-x)E(\mathrm{Te})],
	\label{eq:form_e}
\end{equation}
where $E(\mathrm{Cs}_x\mathrm{Te}_{1-x})$ is the total energy per atom of the relaxed binary crystal, while $E(\mathrm{Cs})$ and $E(\mathrm{Te})$ are the total energies per atom of the most stable Cs and Te elemental phases (Fig.~\ref{fig:pc_workflow}c).
These quantities were obtained from  DFT (\texttt{CP2K} code) using the meta-GGA functional SCAN~\cite{sun+15prl},
improving over the PBE~\cite{perd+96prl} results available at that time in MP and OQMD and delivering enthalpies of formation for the Cs$_5$Te$_3$ and Cs$_2$Te phases in excellent agreement with experiments~\cite{debo-cord97jct}.

The formation energies computed for all the input structures at varying relative Cs and Te content can be visualized with a convex hull connecting the most stable phase for each stoichiometry, including linear combinations of phases with different compositions (Fig.~\ref{fig:pc_workflow}c). 
Adopting a stability criterion (distance from the convex hull) \mbox{$\Delta E_\mathrm{hull}< 0.05~\mathrm{eV/atom}$}, candidate structures below this threshold (Table~\ref{tab:pc_Cs2Te_phases}) are categorized as stable and retained for the subsequent electronic-structure analysis (Fig.~\ref{fig:pc_workflow}d).
Notably, the material at the absolute minimum of the convex hull corresponds to the experimentally known orthorhombic Cs$_2$Te phase (space group $Pnma$)~\cite{sche-boet91}. As illustrated in Fig.~\ref{fig:pc_workflow}d, the band gap of these stable Cs-Te crystals increases almost monotonically with increasing Cs content, culminating in the maximum value obtained for the experimental Cs$_2$Te polymorph. This trend is driven by the gradual filling of the partially occupied Te $p$-orbitals, dominating the frontier states, by electrons donated from the Cs $s$-orbitals~\cite{sass-cocc22jcp}. As the stoichiometry approaches the 2:1 ratio, the depletion of holes in the Te $p$-valence states maximizes the band-gap opening.

\begin{table}[h!]
	\centering
	\caption{\label{tab:pc_Cs2Te_phases} Composition (Comp.), space group, distance to the convex hull ($\Delta E_{hull}$, in eV/atom), band gap ($E_{gap}$, in eV) and database source for the input structure of stable ($\Delta E_{hull}<0.05$~eV/atom) Cs-Te crystals with 2:1 stoichiometry. Structures generated manually from chemically similar compounds are labelled ``m.''.}
    \begin{tabular}{lcccc}
        \textbf{Comp.} & \multicolumn{1}{c}{\textbf{Space group}} & \multicolumn{1}{c}{\mbox{$\Delta E_{hull}$}} & \multicolumn{1}{c}{\mbox{$E_{gap}$}} & \multicolumn{1}{c}{\textbf{Source}} \\
		\hline
		Cs$_8$Te$_4$    & $P1$ [1]           & 0.000 & 2.26 & OQMD \\
		Cs$_8$Te$_4$    & $Pm$ [6]           & 0.000 & 2.25 & OQMD \\
		Cs$_8$Te$_4$    & $P2_1/c$ [14]      & 0.000 & 2.26 & MP m. \\
		Cs$_8$Te$_4$    & $Pnma$ [62]        & 0.000 & 2.25 & MP \\
		\hline
		Cs$_2$Te    & $R\bar{3}m$ [166]  & 0.002 & 2.14 & OQMD \\
		Cs$_2$Te    & $Fm\bar{3}m$ [225] & 0.002 & 2.14 & OQMD \\
		\hline
		Cs$_{18}$Te$_9$ & $R3$ [146]         & 0.004 & 1.96 & OQMD m. \\
		\hline
		Cs$_4$Te$_2$    & $Cmc2_1$ [36]        & 0.018 & 2.16 & MP m. \\
		Cs$_4$Te$_2$    & $Fdd2$ [43]        & 0.018 & 2.18 & MP m. \\
		Cs$_4$Te$_2$    & $P6_3/mmc$ [194]   & 0.018 & 2.22 & MP m. \\
    \end{tabular}
\end{table}

Since surface composition and morphology play a critical role in photoemission, we extended \texttt{aim$^2$dat} implementing the \texttt{SurfaceGenerator} class. Corresponding routines enable the construction of surface facets, a unique feature that is not available in any other high-throughput screening package at the time of this development~\cite{sass-cocc24npjcm}. To test the functionality of this new class, we generated low-Miller-index surfaces of Cs$_2$Te, considering all possible terminations of each stable polymorph identified in the high-throughput screening of bulk structures~\cite{sass-cocc22jcp}.

\begin{figure*}
    \begin{center}
		\includegraphics[scale=0.75]{surf_generator_v2.pdf}

		\caption{\label{fig:pc_surf_gen} Schematic representation of the \texttt{SurfaceGenerator} architecture. The workflow illustrates the transformation of bulk structures into oriented slabs, with white boxes representing input metadata, shaded boxes indicating automated execution steps, and filled boxes denoting the finalized \texttt{SurfaceData} output.}
	\end{center}
\end{figure*}

Surface generation is carried out in three steps visualized in Fig.~\ref{fig:pc_surf_gen}:
\begin{enumerate}
	\item \textbf{Unit Cell Standardization:} The conventional unit cell is determined using the \texttt{spglib}~\cite{spglib} package to ensure a consistent starting point.
	\item \textbf{Initial Slab Construction:} A trial slab with the targeted Miller indices is generated using the \textit{surface} function of the \texttt{ase}~\cite{ase} package.
	\item \textbf{Modular Decomposition:} The toolkit detects the repeating unit cell inside the slab (including translation vectors), along with the unit cells for the bottom and top terminations.
\end{enumerate}
By storing this information in the \texttt{SurfaceData} object, \texttt{aim$^2$dat} can construct slabs of arbitrary thickness on-the-fly by vertically stacking these fundamental building blocks. This modularity allows the \texttt{SurfaceOptWorkChain} class to efficiently perform thickness-convergence tests and relax the atomic positions. Subsequent analysis of the electronic properties of the converged and relaxed slab, including band structures, projected density of states (pDOS), and partial charges, is seamlessly handled by the \texttt{BandStructureWorkChain}, \texttt{PDOSWorkChain}, and \texttt{PartialChargesWorkchain} routines, respectively.
To ensure the numerical stability of these surface calculations and ensure adequate accuracy, we employed the r$^2$SCAN functional~\cite{furn+20jpcl} for the DFT calculations~\cite{sass-cocc24npjcm}.

Surface stability was determined for each slab by computing the surface energy as a function of the relative Cs content at a given chemical potential (Fig.~\ref{fig:pc_workflow}e).
This analysis revealed three different regimes based on surface stoichiometry: (i) Cs-rich surfaces (negative slope in Fig.~\ref{fig:pc_workflow}e), exhibit a metallic behavior and a reduced ionization potential (Fig.~\ref{fig:pc_workflow}f); (ii) Stoichiometric surfaces (horizontal lines in Fig.~\ref{fig:pc_workflow}e), which maintain bulk stoichiometry, serve as a baseline for electronic properties; (iii) Te-rich surfaces (positive slope in Fig.~\ref{fig:pc_workflow}e), characterized by large band gaps and increased ionization potentials, that eventually plateau at the bulk value (Fig.~\ref{fig:pc_workflow}f).
These stoichiometric variations, uncovered by the automated screening implemented in \texttt{aim$^2$dat}, ultimately govern photocathode performance. Specifically, the accumulation of excess Cs atoms at the outermost layers acts as the primary driver for lowering the emission barrier, although simultaneously quenching the band gap~\cite{sass-cocc24npjcm}.

\subsection{Automated Compositional Tuning of Metal-Organic Frameworks}\label{example:mofs}

\begin{figure*}
	\centering
    \includegraphics[width=\linewidth]{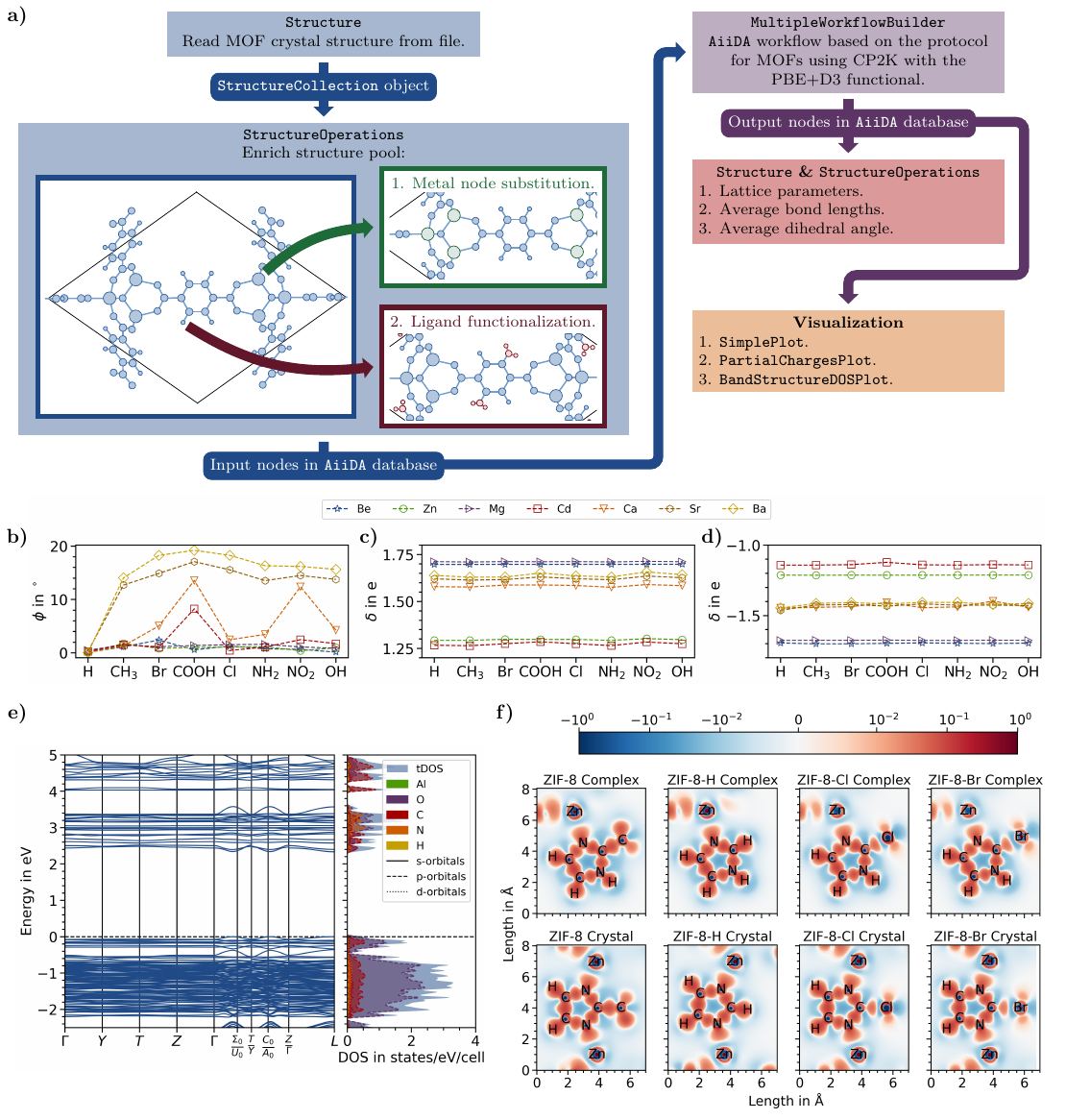}

    \caption{\label{fig:mof} Automated structural and electronic analysis of MOF and ZIF derivatives. (a) Schematic of the \texttt{aim$^2$dat} construction workflow for functionalized MOF derivatives, identifying individual tasks (rectangles) and their respective data outputs (rounded rectangles). (b) Comparative analysis of the average rotation angle ($\phi$) of the BDC linker across the MOF-5 derivative ensemble. (c–d) Distribution of partial charges ($\delta$) for the metal species (M) and the central oxo-cluster oxygen (O1), highlighting the impact of functionalization on the local electrostatic environment. (e) Electronic band structure and pDOS for Al$_2$(BDC)$_3$, with the valence band maximum (VBM) referenced to 0 eV. (f) Electron deformation density maps for ZIF-8 derivatives, illustrating the chemical bonding environment within the heterocyclic ring planes. Data sourced from Refs.~\cite{mof_mof5derivatives_zenodo, mof_scmof_zenodo, mof_zif_8_zenodo}.}
\end{figure*}

Metal–organic frameworks are a versatile class of porous materials characterized by a periodic network of metal nodes coordinated by molecular linkers. The primary challenge in modelling MOFs lies in this hybrid nature. Their porous character leads to significantly larger unit cells compared to those of conventional inorganic crystals, and their properties are governed by a delicate interplay between the local chemistry of the organic ligands and the global symmetry of the framework~\cite{form+24afm}. Simultaneously, this complexity offers a unique platform for structural and chemical tunability, which has been largely exploited for specific applications, such as, for instance, gas storage and separation~\cite{suh+12cr,alti+23mte}, as well as photocatalysis and photoelectric conversion~\cite{nava+23cr}.

The large configurational space that naturally characterizes MOFs can be efficiently tamed via high-throughput screening~\cite{colo-snur14csr}.
By targeting prototypical MOF structures like MOF-5 and scandium terephthalate (Sc$_2$(BDC)$_3$), we applied \texttt{aim$^2$dat} to exchange the metal node and/or functionalize their linkers~\cite{sass+24ic,edza+24jcp}. In the case of scandium terephthalate, this optimization is further motivated by the need to replace a rare-earth element like Sc with a more abundant species without compromising the intrinsic properties of the MOF~\cite{bars+18rcr}. Since both MOF-5 and Sc$_2$(BDC)$_3$ contain the same ligand (benzenedicarboxylic acid, BDC), we functionalized both with CH$_3$, NO$_2$, Cl, Br, NH$_2$, OH, and COOH groups.

The construction of the functionalized MOF derivatives followed a hierarchical path, starting with the ingestion of the primary parent structure into the \texttt{StructureCollection} class (Fig.~\ref{fig:mof}a). The structure was modified in two main stages (see second box in Fig.~\ref{fig:mof}a): isomorphous metal substitution and functional group grafting. First,  the metal nodes are replaced using the \texttt{StructureOperations} class. To provide a reliable starting point for DFT relaxation, the lattice parameters are scaled based on the ratio of the covalent radii of the substituted elements. This geometric approximation ensured physically reasonable initial configurations, thereby significantly reducing the \textit{ab initio} computational costs. In the second stage,  functional groups are bound to specific sites within the ligand.
While the host atoms were identified manually in our first studies~\cite{sass+24ic,edza+24jcp}, \texttt{aim$^2$dat} automates guest positioning using sophisticated geometric heuristics. Specifically, the toolkit calculates the local environment of the host atom and generates a placement vector designed to maximize the distance between the guest molecule and the neighbouring atoms, simultaneously minimizing steric hindrance. Furthermore, \texttt{aim$^2$dat} supports multicentre coordination, allowing guest molecules to be positioned relative to a cluster of atoms -- an essential feature for accurately modelling complex binding sites within the pores.

Most aspects of the workflow for MOFs share the same logic elaborated in Sec.~\ref{example:pc}. In addition, the protocol \texttt{cp2k-crystal-MOFs-standard\_v2.0} introduces two specialized tasks handling the complexity of these hybrid crystals.
First, to address the structural distortions often introduced by large-scale metal substitution or functional group grafting, the protocol utilizes a hierarchical optimization strategy. A pre-relaxation of the MOF is performed with the \texttt{CellOptWorkChain} routine with reduced plane-wave cutoffs and smaller basis sets. This step significantly boosts the convergence of the subsequent optimization, where the electronic and structural properties are refined with production-level settings. Second, the \texttt{PartialChargesWorkChain} class facilitates the extraction of partial charges, which are critical for understanding the electrostatic environment within the pores. \texttt{aim$^2$dat} supports both the DDEC6~\cite{manz-lima16rscadv} and Bader~\cite{bade-zou92cpl} methods.
For flexible post-processing, users can extract individual atomic charges or generate a localized cluster of charges, enabling the investigation of specific active sites or the global electrostatic potential surface of the MOF.

The results stored in the \texttt{AiiDA} output nodes are seamlessly re-integrated into the \texttt{StructureCollection} class, enabling immediate geometric analysis. To characterize the local atomic environment, \texttt{aim$^2$dat} calculates average bond lengths and coordination by evaluating the neighbour list of each atomic site. The calculation of dihedral angles allows for detailed conformational analysis; specifically, users can either specify atomic indices or leverage the software's ability to identify symmetry-equivalent sites to automate the assessment of torsional angles across the entire framework. These analytical features, combined with the built-in visualization tools, provide a comprehensive platform for quantifying how compositional tuning impacts the structural integrity and internal geometry of complex hybrid materials.

As an example, Fig.~\ref{fig:mof}b illustrates the distortion angle formed at the center of the functionalized BDC linker in several MOF-5 derivatives~\cite{edza+24jcp}.
The rotation angle was evaluated with respect to the pristine structure by identifying equivalent atoms, including the central oxygen of the oxo-cluster and the carbon atoms linking the phenyl ring to the oxo-cluster.
For each derivative in the structural ensemble, the individual rotation angles were computed and averaged to yield the representative conformational state.

The analysis indicates that the functionalization with smaller metal nodes has a negligible effect on structural distortion, as functional groups retain sufficient steric space, avoiding significant interactions with the framework. Conversely, larger metal nodes prefer coordination numbers larger than four, triggering a rotation of the BDC linker to facilitate multi-site coordination.
This behavior is particularly evident in Cd and Ca derivatives, where the partially negative carboxylic group interacts strongly with the electropositive metal node. Furthermore, the larger ionic radius of calcium promotes additional interactions with nitrogen dioxide groups, leading to further lattice distortion. Each functional group paired with these larger metal nodes exerts a repulsive force against the oxo-cluster, resulting in the observed structural tilt.
In addition, calcium, a larger metal node than cadmium, interacts with nitrogen dioxide and leads to the distortion.
Each functional group paired with the larger metal nodes interacts with the oxo-cluster, and results in a repulsive interaction leading to the distortion.
These results are visualized with the \texttt{SimplePlot} class, which abstracts the complexity of data formatting. As shown in Figure~\ref{fig:mof}b, the plotting interface allows for the direct mapping of functional groups to the $x$-axis, overlaying results for various metal-substituted MOFs for immediate comparison. This visualization pipeline enables the quick identification of structural distortions in specific frameworks.

The \texttt{PartialChargesPlot} class enables direct data import from either the AiiDA node or external sources. The plotting interface automatically parses the structural dataset to generate element-specific subsets, computing statistical averages for atomic charges across the ensemble. As shown in Fig.~\ref{fig:mof}c-d, this class then renders these results as comparative subplots. This automated aggregation is particularly useful for identifying trends; for example, when multiple structures share a common metallic node, the dataset can be aggregated to compare the effects of different functional groups on the local electrostatic environment of the MOF.

In general, the inclusion of functional groups exerts only a minor influence on the charges at the metal nodes (Fig.~\ref{fig:mof}d) and the central oxygen of the oxo-cluster (Fig.~\ref{fig:mof}e).
This is attributed to the relative distance from the functionalization site, where Coulomb interactions are significantly attenuated compared to the binding forces of neighboring atoms.
The oxygen charge effectively mirrors the charge of the metal node, indicating consistent electrons shift towards the more electronegative oxygen centers.

The distribution of these charges into two distinct categories aligns with the underlying electronic structure. The charge of the transition metals (e.g., Cd and Zn) is effectively shielded by their filled $d$-shell, resulting in a lower effective charge compared to that of the alkaline earth elements.
An additional split is observed within the alkaline earth metals between smaller (Be and Mg) and larger metal nodes (Ca, Sr and Ba). This divergence arises from the shorter bond lengths in the smaller analogs, which leads to enhanced orbital overlap and stronger interactions with the neighboring oxygen atoms~\cite{edza+24jcp}.

The \texttt{BandStructureDOSPlot} class is built upon two independent components, \texttt{BandStructurePlot} and \texttt{DOSPlot}, which can be utilized separately or combined for a comprehensive overview of the electronic structure.
The datasets are ingested from the \texttt{AiiDA} output nodes, extracting the band structure along paths connecting high-symmetry points in \textbf{k}-space predefined in the workflow. For the pDOS visualization, the toolkit processes each atom individually and resolves the contributions by angular momentum ($l$-quantum number). This orbital-resolved approach provides in-depth insights into how specific atomic orbitals contribute to the frontier states. The total density of states is then computed as the aggregate of these projected contributions across the relevant energy range. By automating the alignment of the Fermi level and the mapping of the \textbf{k}-path, these classes allow for the rapid, high-throughput generation of publication-ready electronic structure diagrams, as illustrated in Figure~\ref{fig:mof}e.
The observation of the band structure indicate localized states due to the flat bands in the valance region along $\Gamma-Y-T-Z-\Gamma$, whereas a more significant dispersion is noticed in other areas.
The relationship between band dispersion and the orientation of the linker molecules within the unit cell can be established: in the (reciprocal) directions that align with the carbon conjugation of the linker, there is an increase in electron mobility, resulting in more dispersive bands.
This explanation is corroborated by the analysis of the projected density of states (pDOS), which indicates a link between bands that display substantial dispersion and states predominantly influenced by C p-orbitals, particularly in the conduction region.
The electronic levels closest to the frontier exhibit carbon–oxygen hybridization, indicating the contribution of the CO$_2$ at the connection between the BDC linkers and the metal nodes
~\cite{edza+24jcp}.

Zeolitic imidazolate frameworks (ZIFs) form a special class of MOFs in which imidazole-based ligands bridge metal centers to form crystalline structures with topological arrangements similar to zeolites.
ZIF-8, a prominent representative of this class of materials, exhibits extraordinary structural changes upon external stimuli~\cite{iaco-maur21acsami}, stemming from its peculiar electronic charge distribution and chemical bonds.
To study the impact of ligand substitution on the charge distribution in ZIF-8 derivatives~\cite{edza+23jpcc}, \texttt{aim$^2$dat} was employed to construct the input structures for the DFT calculations by substituting passivating hydrogen atoms with functional groups.
The routines implemented to calculate the molecular planes of the heterocyclic rings combined with the interface to the \texttt{critic2} code facilitated the calculation of the electron deformation density (Fig.~\ref{fig:mof}f).

The distribution of the deformation density across the carbon-based backbone reveals a continuous electron sharing characteristic of covalent bonds, which appears largely unaffected by the degree of periodicity (whether in bulk crystals or isolated complexes). In the presence of CH$_3$ and H terminations, the functional groups exhibit a strong covalent attachment to the backbone. Conversely, the halogen atoms are linked to the inner network through highly polar covalent bonds, as evidenced by the localized depletion of electron density (bluish areas in Fig.~\ref{fig:mof}f) in the bonding region. Similarly, the pronounced ionic character of the Zn–N bond is observable across all middle and lower panels of Fig.~\ref{fig:mof}f, underscoring the complex bonding in ZIF-8.

\subsection{Machine-Learning-Assisted Crystal Structure Prediction}\label{example:ml}

\begin{figure*}
	\begin{center}
		\includegraphics[width=\linewidth]{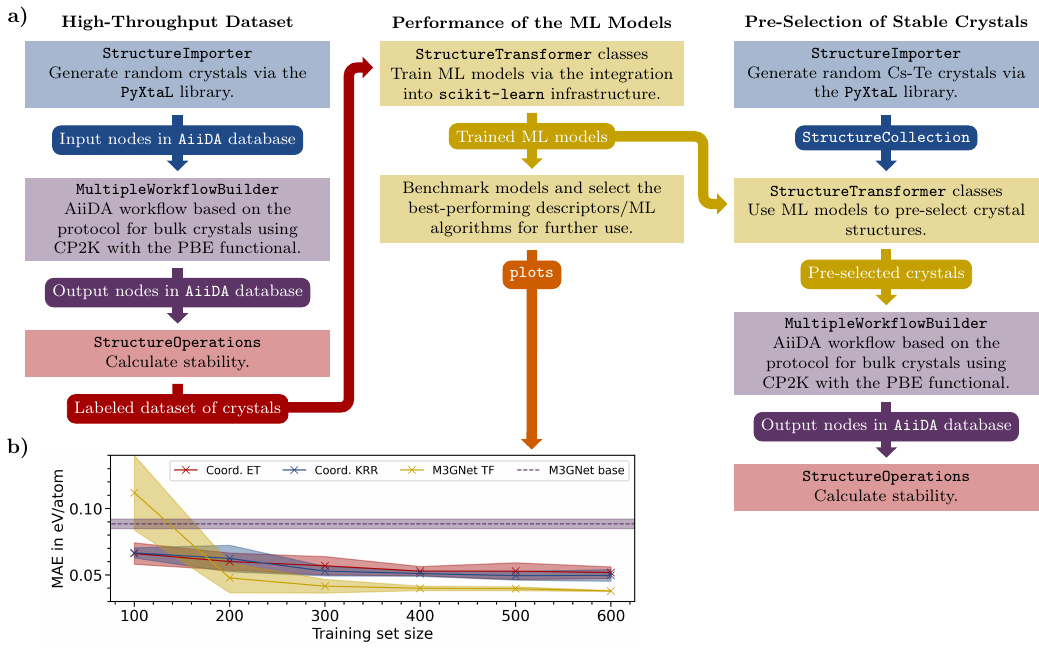}

		\caption{\label{fig:ml_workflow} (a) Schematic overview of the high-throughput pipeline for ML model training, illustrating the sequence of tasks (rectangles) and their output (rounded boxes). (b) Predictive performance of the trained model, showing the Mean absolute error (MAE) averaged over three random data splits. The shaded region denotes the standard deviation, representing the model's stability across different training subsets. Data sourced from Ref~\cite{ml_zenodo}.}
	\end{center}
\end{figure*}

Crystal structure prediction (CSP) aims to identify the ground-state structure of a material solely from its stoichiometry.
Since atomic positions and lattice vectors are prerequisites of any \textit{ab initio} calculation, CSP is a transformative technique for discovering new phases and crystal structures before their synthesis.
Over the past several decades, various CSP methods have emerged, ranging from fixed stoichiometry searches~\cite{goed10, yama+18prm} to explorations of binary and ternary crystalline materials~\cite{kvas+20cm}.
We developed and applied a novel CSP scheme to Cs-Te crystals~\cite{sass-cocc25ats}, leveraging the ability of \texttt{aim$^2$dat} to manage large ensembles of randomly generated crystal structures. Combining high-throughput DFT calculations with supervised ML, we efficiently estimated structural stability and ranked the crystals accordingly.

The workflow sketched in Fig.~\ref{fig:ml_workflow}a demonstrates how \texttt{aim$^2$dat} acts as the orchestrator for a multi-stage predictive pipeline.
The process begins with the generation of a vast structural ensemble featuring varying Cs concentrations. This is achieved using the \texttt{StructureImporter} class, providing the input structural pool.
High-throughput DFT calculations are performed to minimize interatomic forces and internal pressure. The energetic stability of each crystal structure is subsequently quantified by the formation energy ($E_\mathrm{f}$) and the distance from the convex hull ($\Delta E_\mathrm{Hull}$).
At this stage, 34.2\,\% of the initial structural pool failed reaching convergence, primarily due to unfavorable coordination environments in the randomly generated structures that led to numerical instabilities in the DFT iteration~\cite{sass-cocc25ats}. Despite these challenges, the automated error-handling and filtering routines of \texttt{aim$^2$dat} ensured a final dataset of 1073 successfully relaxed crystals. These served as a basis for the subsequent ML training phase~\cite{sass-cocc25ats}.

To train and validate the ML models, the dataset was partitioned into multiple training datasets, ranging from 100 to 600 crystal structures, with a consistent test set including the remaining 473 structures~\cite{sass-cocc25ats}. We focused on two main approaches.
The first one is based on the \textit{Coord.} descriptor, implemented in \texttt{aim$^2$dat}, which encodes the local coordination environment of the atomic sites  into a global feature vector by calculating the average, minimum, and maximum of local quantities such as the number of neighboring atoms and their respective distances.
It is combined with the extremely randomized trees (\textit{ET}) or kernel ridge regression (\textit{KRR}) algorithms available from the \texttt{scikit-learn} infrastructure to predict energetic stability.
The second scheme leverages the \textit{M3GNet} model, a state-of-the-art graph neural network architecture implemented in the \texttt{matgl} package~\cite{ko+25npjcm}, which was pre-trained on over 60,000 crystals.
We compared the \textit{M3GNet base} model, merely relying on the pre-training, against the \textit{M3GNet TF} variant that has undergone transfer learning on our specific Cs-Te structural dataset.

The mean absolute error (MAE) defined as
\begin{equation}
    \text{MAE} = \frac{1}{n}\sum |y_i - \hat{y}_i|,
\end{equation}
where $n$ is the number of materials in the evaluation set, $y_i$ represents the reference value calculated from DFT, and $\hat{y}_i$ is the value predicted by the model, is used as the primary metric to compare the accuracy of the selected ML models (Fig.~\ref{fig:ml_workflow}b)
Since the \textit{Coord. ET} and the \textit{Coord. KRR} models use the same descriptor, they exhibit comparable performance, with a prediction error of approximately 0.05~eV/atom when trained on the largest subsets.
The transfer-learned \textit{M3GNet} model achieves the highest accuracy, with an MAE dropping below 0.04~eV/atom.
This suggests that the combination of a high-capacity graph neural network with domain-specific fine-tuning effectively captures the complex potential energy surface of the Cs-Te system. On the other hand, the \textit{M3GNet base} variant produces the highest error, as expected, considering that this model was not trained on this specific dataset. These results highlight the necessity of dataset-specific training, facilitated by the automated data generation in \texttt{aim$^2$dat}, to achieve chemically accurate stability predictions in novel structural domains.

\begin{table}
	\centering
	\caption{Mean energetic stability ($\Delta E_\mathrm{Hull}$) and MAE of the predicted stability in eV/atom; success rate (S.R., in \%) of the high-throughput workflow and relative amount of stable elemental Te / mixed / elemental Cs crystal structures (in \%). The results delivered by the best-performing models are highlighted in bold.}
	\label{tab:ml_results}
	\begin{tabular}{lcccc}
		& \textbf{$\Delta E_\mathrm{hull}$} & \textbf{MAE} & \textbf{S.R.} & \textbf{Stable crystals} \\
		\hline
		Initial pool & 0.212 & \multicolumn{1}{c}{--} & 65.8 & 0.0/2.1/1.3 \\
		\textit{Coord. ET} & 0.178 & \textbf{0.081} & \textbf{90.0} & \textbf{0.0/6.0/5.0} \\
		\textit{Coord. KRR} & 0.164 & 0.101 & 80.0 & 0.0/5.0/5.0 \\
		\textit{M3GNet TF}    & 0.195 & 0.110 & \textbf{90.0} & 1.0/5.0/1.0 \\
		\textit{M3GNet base}  & 0.193 & 0.156 & 77.0 & 0.0/4.0/3.0 \\
	\end{tabular}
\end{table}

In the final stage of the CSP pipeline, a second structure pool, containing 15,000 candidate crystals, was generated to mirror the diversity of the initial set.
To circumvent the prohibitive cost of ``brute-force'' DFT optimizations for a dataset of this size, the trained ML models were deployed as a high-speed pre-screening filter.
To further refine the structure selection, \texttt{aim$^2$dat} implements a symmetry-constrained grid search method that systematically varies lattice vectors and angles. This approach identifies the most energetically favorable variant for each configuration while preserving the underlying structural motifs.
Following this pre-screening, the top 100 candidates were chosen and validated via DFT calculations (see Tab.~\ref{tab:ml_results}).
In contrast to the learning curves displayed in Fig.~\ref{fig:ml_workflow}b, the \textit{Coord.} models show the best performance, identifying the largest fraction of stable crystals.
This outcome suggests that the coordination-based descriptors are particularly robust when navigating previously unseen regions of the configurational space. Ultimately, this methodology offers a powerful and scalable pathway for discovering stable phases in experimentally challenging domains, such as more complex ternary landscapes of multi-alkali antimonides.

\begin{figure*}
    \begin{center}
		\includegraphics[scale=0.8]{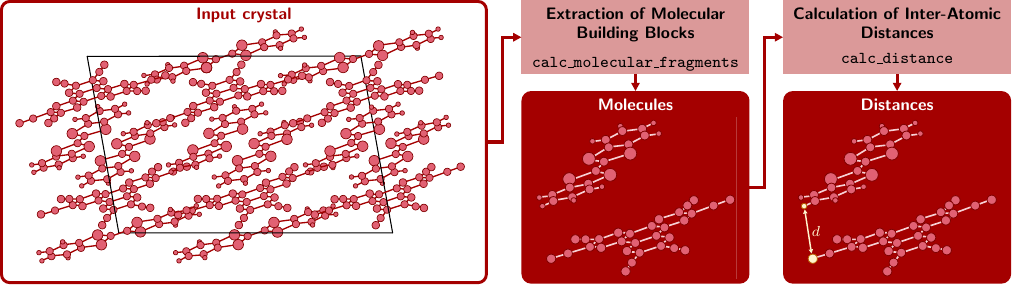}

		\caption{\label{fig:fragmentation} Schematic overview of the \texttt{aim$^2$dat} workflow for the unsupervised identification of molecular fragments using \texttt{calc\_molecular\_fragments} function, followed by the automated calculation of inter-fragment proximity minimal distances via \texttt{calc\_distance}.}
	\end{center}
\end{figure*}

\subsection{Other use-cases}\label{example:misc}

Beyond studies planned to include high-throughput screening and workflow automation, \texttt{aim$^2$dat} was used as an efficient pre- and post-processing tool for structural analysis, visualization, and to unravel correlations in complex datasets.

The first notable example concerns scandium fluorides and oxides, which are classes of technologically relevant materials~\cite{iver+01jecs,liu+10tsf,dori+18,wu+24am}, characterized by a pronounced d polymorphism. To identify the signatures of (meta)stable crystal structures in experimental X-ray absorption spectra~\cite{mach+23ic,duar+25rsca}, \texttt{aim$^2$dat} was used to calculate inter-atomic distances and quantify the similarity of different crystalline phases using the F-Fingerprint metric~\cite{ogan-vall09jcp}.

\begin{figure*}
    \centering
    \includegraphics[width=\linewidth]{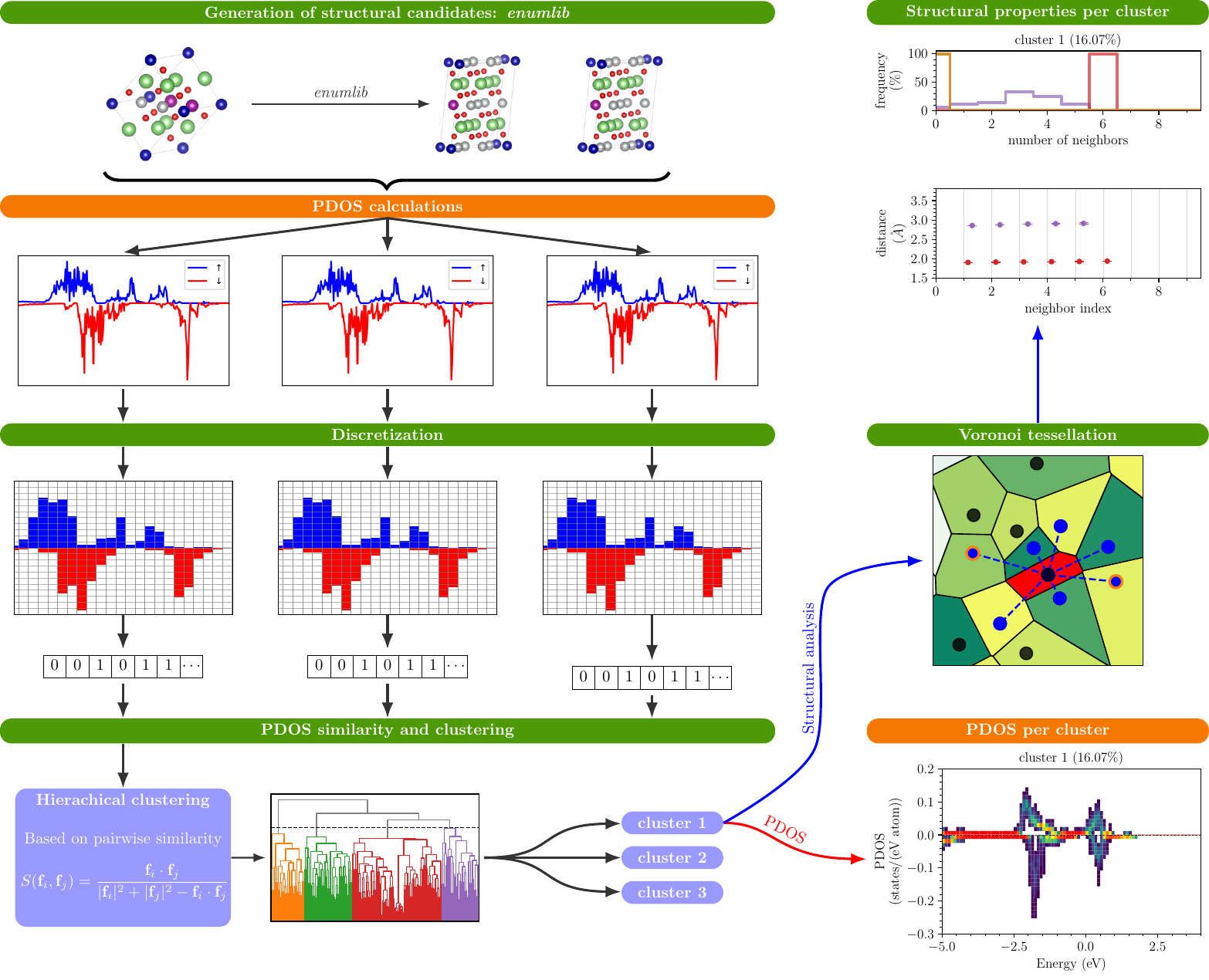}
    \caption{
        Schematic representation of the high-throughput workflow used to correlate local coordination environments with the electronic properties of battery cathode materials~\cite{reen+25sr}. Modules natively implemented in \texttt{aim$^2$dat} are highlighted in green, while external library dependencies are shown in orange. The workflow starts with the automated generation of structural candidates from a parent lattice.  Subsequently, the calculated pDOS is discretized to facilitate hierarchical clustering based on spectral similarity. In the final stage, site-resolved electronic descriptors are combined with structural features, including coordination numbers and the inter-atomic distances derived from Voronoi-tessellation, to provide an aggregated profile for each cluster.
    }
    \label{fig:reents-NCM-paper}
\end{figure*}

Organic cocrystals (OCCs) offer tuneable electronic and optical properties based on molecular components and intermolecular interactions~\cite{liu+23cem,anha+26afm}. When they are formed by donor-acceptor pairs, exhibiting large charge-transfer, their properties are especially sensitive to the crystal packing structure and the donor:acceptor ratio. In a recent study dedicated to unravel the electronic and properties of ternary OCCs~\cite{vale+26cm}, \texttt{aim$^2$dat} was used to calculate inter-atomic distances in different molecules in the crystal.
The adopted workflow is organized in three steps (Fig.~\ref{fig:fragmentation}).
First, the crystal is loaded in the \texttt{Structure} class.
Second, the analysis function \texttt{calc\_molecular\_fragments} is deployed to find the molecules in the crystal.
The function constructs a supercell, ensuring the correct identification of molecules that surpass the periodic boundaries of the unitcell.
By constructing a graph based on the neighbours of each site, the algorithm groups all atoms into the molecular components.
Third, the distances between each atom pair belonging to different molecules is calculated, and the minimum distance is retrieved.

The capabilities of \texttt{aim$^2$dat} were further leveraged to investigate the complex mapping between local structural environments and their corresponding electronic signatures in the pDOS of ternary transition-metal oxides for battery materials~\cite{reen+25sr}. Specifically, the workflow automation and the analysis tools were used to identify structural and electronic descriptors that provided a mechanistic interpretation of X-ray spectroscopy data.
To achieve this, \texttt{aim$^2$dat} provides post-processing modules that discretize continuous spectral data and pDOS curves into vectorized representations.
These electronic fingerprints are then paired with geometric descriptors—such as local coordination numbers and bond-distance distributions derived from Voronoi tessellation, enabling automatic grouping of structures with similar optoelectronic and local environments.
The workflow illustrated in Fig.~\ref{fig:reents-NCM-paper} starts with the generation of structural candidates based on a parent structure. This enumeration is performed through the \texttt{enumlib} library~\cite{hart+12cms} integrated into the workflows of \texttt{aim$^2$dat} implementing an \texttt{AiiDA} \texttt{CalcJob} element.

After structural relaxation of the generated structural candidates performed with \texttt{Quantum ESPRESSO}~\cite{gian+09jpcm,gian+17jpcm} interfaced to \texttt{AiiDA}~\cite{web:aiida-quantumespresso}, the  pDOS is calculated and subsequently  discretized to be converted into a binary vector~\cite{kuba+22sd}.
This representation is used to calculate pairwise similarities for hierarchical clustering to group sites exhibiting similar electronic-structure features. By creating a 2D-histogram over all sites in a cluster, the pDOS per cluster is inspected using Voronoi-tessellation implemented in the \texttt{StructureOperations} class to determine the local environment and coordination of each site and, subsequently, the distances between the atom of interest and its neighbors.

An analogous strategy can be applied to aggregate structural information across all sites per cluster and identify a relation between the average electronic and structural features and trends within each cluster. In the example with transition-metal oxides~\cite{reen+25sr}, the gathered information is used to fit the best matching combination of the electronic fingerprints per cluster with respect to the experimental X-ray spectroscopy data to infer information about the local structural features in the material. 

\section{Conclusions}\label{sec:conclusions}

In summary, we have presented \texttt{aim$^2$dat}, an open-source \texttt{Python} ecosystem designed to serve as a critical bridge between high-throughput computational materials science and modern data science. By providing a unified and extensible user interface, this package seamlessly integrates structural generation, automated simulation workflows, and advanced data visualization with the predictive power of machine learning.

Throughout the presented use cases, we have demonstrated that \texttt{aim$^2$dat} is more than a simple collection of scripts: it is a versatile framework capable of navigating diverse chemical spaces. Whether managing the screening of alkali-based semiconductors, resolving the complex coordination environments of functionalized metal-organic frameworks, or mapping local structural motifs to spectroscopic signatures, this library provides a consistent logic for data curation and analysis. The decoupled architecture of its visualization and processing modules specifically addresses the bottleneck of high-throughput studies, allowing researchers to rapidly iterate from raw simulation outputs to mechanistic physical insights.

As the field of materials science increasingly shifts toward AI-driven discovery, tools like \texttt{aim$^2$dat} that ensure data interoperability and workflow reproducibility will become indispensable.
While the current workflows focus primarily on ground-state properties, the modular framework of the library is uniquely positioned to accommodate future extensions, such as the integration of finite-temperature effects or entropic stabilization via automated phonon and molecular dynamics pipelines.
All modules are fully open-source and hosted on \texttt{GitHub}~\cite{repository} and accompanied by extensive documentation and examples~\cite{docs} for frictionless access and broad dissemination. We anticipate that \texttt{aim$^2$dat} will empower researchers to scale their computational efforts, moving beyond the study of individual systems toward the broad, data-centric exploration of the materials genome.

\section*{Acknowledgements}

The authors are grateful to Hilde Bellersen, Daniel Guo, Fabiana Machado Ferreira De Araujo, Julia Santana Andreo, Richard Schier, and Ana Maria Valencia for extensive feedback on the software.

\section*{Funding}

This work was funded by the German Federal Ministry of Education and Research (Professorinnenprogramm III), by the Lower Saxony State (Professorinnen f\"ur Niedersachsen and SMART), and by the German Research Foundation (Deutsche Forschungsgemeinschaft, DFG) project No. 561059489 (Walter-Benjamin Fellowship, H.-D.S), 490940284 (C.C.), and 398816777 (CRC 1375 ``NOA", subproject A8, C.C.). J.E. appreciates financial support from the Nagelschneider Stiftung.

\section*{Author contributions}

\textbf{H.-D.S.} Conceptualization, Software, Methodology, Writing – original draft; \textbf{J.E.} Software, Methodology, Writing – original draft; \textbf{T.R.} Software, Methodology, Writing – original draft; \textbf{C.C.} Conceptualization, Funding acquisition, Resources, Supervision, Writing – review \& editing

\bibliography{bib}

\end{document}